\documentclass[twocolumn]{aastex631}

\shortauthors{Sun et al.}
\graphicspath{{./}{figures/}}
\begin{document}

\title{A 3D Chemodynamical Census of Inner-Galaxy Metal-poor Giants to [Fe/H]$\sim-3.5$}

\newcommand{\KIAA}{\affiliation{Kavli Institute for Astronomy and
Astrophysics, Peking University, Beijing 100871, People's Republic of China}}
\newcommand{\DoA}{\affiliation{Department of Astronomy, School of Physics,
Peking University, Beijing 100871, People's Republic of China}}
\newcommand{\UCAS}{\affiliation{School of Astronomy and Space Science, University of Chinese Academy of Sciences, Beijing 100049, People's Republic of China}}
\newcommand{\NAOC}{\affiliation{National Astronomical Observatories, Chinese Academy of Sciences, Beijing 100101, People's Republic of China}}
\newcommand{\SJTU}{\affiliation{Department of Astronomy, School of Physics and Astronomy, Shanghai Jiao Tong University, 800 Dongchuan Road, Shanghai 200240, People's Republic of China}}
\newcommand{\IFAA}{\affiliation{Institute for Frontiers in Astronomy and Astrophysics, Beijing Normal University, Beijing 102206, People's Republic of China}}
\newcommand{\KLDM}{\affiliation{State Key Laboratory of Dark Matter Physics, School of Physics and Astronomy, Shanghai Jiao Tong University, Shanghai 200240, People's Republic of China}} 
\newcommand{\ND}{\affiliation{Department of Physics and Astronomy, University of Notre Dame, Notre Dame, IN 46556, USA}}
\newcommand{\JINA}{\affiliation{Joint Institute for Nuclear Astrophysics -- Center for the Evolution of the Elements (JINA-CEE), USA}}

\correspondingauthor{Yang Huang, Fanzhou Jiang, Huawei Zhang}
\email{huangyang@ucas.ac.cn, fangzhou.jiang@pku.edu.cn, zhanghw@pku.edu.cn}

\author[0000-0002-9796-1507]{Shenglan Sun}\DoA\KIAA

\author[0000-0003-3250-2876]{Yang Huang}\UCAS\NAOC

\author[0000-0001-6115-0633]{Fangzhou Jiang}\KIAA

\author[0000-0002-7727-1699]{Huawei Zhang}\DoA\KIAA

\author[0000-0002-0642-5689]{Xiang-Xiang Xue}\NAOC\IFAA

\author[0000-0003-4573-6233]{Timothy C. Beers}\ND\JINA

\author[0000-0001-6655-854X]{Chengye Cao}\SJTU

\author{Qikang Feng}\DoA\KIAA

\author[0009-0008-1319-1084]{Ruizhi Zhang}\NAOC\UCAS

\author{Haiyang Xing}\DoA\KIAA

\author[0000-0002-7662-5475]{João A. S. Amarante}\SJTU\KLDM



\begin{abstract}
The earliest assembly of the Milky Way remains poorly understood, yet the spatial, chemical, and kinematic properties of its most metal-poor stars provide a unique fossil record of its proto-Galaxy phase. Understanding how this ancient component formed is essential for linking near-field Galactic archaeology to high-redshift galaxy evolution.
We construct the currently largest 3D map of inner-Galaxy metal-poor giants by combining several narrow/medium-band photometric surveys, reaching metallicities down to $\mathrm{[Fe/H]}\sim-3.5$. Our final sample contains 5,095,676 giants, including 1,717,610 stars with $\mathrm{[Fe/H]}<-1$.
Across $-4\le \mathrm{[Fe/H]}<-1$, the density distribution reveals a centrally concentrated, flattened spheroidal component extending to $r_{\rm gc}\sim15$ kpc, together with a prominent overdensity near $X\sim-5$ kpc that is dominated by metal-poor stars on disklike orbits, with a kinematically hot background also present. The selection-function-corrected metallicity distribution function shows a distinct, very metal-poor component around $\mathrm{[Fe/H]}\sim-2.7$ that becomes most prominent at 1~$<r_{\rm gc}<$~3 kpc. Stars with $-3.5\lesssim\mathrm{[Fe/H]}\lesssim-1.4$ exhibit weak net rotation and low rotational support within $r_{\rm gc}<15$ kpc. Finally, we briefly note that the centrally enhanced very metal-poor component could be qualitatively consistent with one or more early dissipative build-up episodes (e.g., high-$z$ compaction/``blue-nugget'' phases) as one possible interpretation.
\end{abstract}



\section{Introduction} 
%第一段（最大动机）：研究MW形成历史，用最古老恒星。
%第二段（把古老转换成贫金属）：直接测年龄不容易，可以用金属丰度。realm那篇第一段，贫金属星在星系中心的原因。列举几篇观测结果，说明贫金属星确实在星系中心。
%第三段（星系中心贫金属星研究现状）：thanks to 一堆巡天，目前对内晕的理解是innermost population + GES + smaller substructures，对后两个已经研究over decades，但前者more elusive，被称作proto Galaxy。Fortunately,最近一些工作已经开始揭示这一成分了。
%第四段（现状详细版+当前限制）：列举几篇，对proto Galaxy的空间分布、运动学、化学都以有限的样本和金属丰度下限进行了刻画，但是没有拓展到更低的金属丰度。thanks to 中窄带测光巡天，有了大规模金属丰度和距离测量。标榜我们这篇低到了-3.5。
%第五段（缺乏proto Galaxy形成机制的研究）：列举cold filament (BK22), starburst due to merger (chen2025)，引出blue nugget，先说一嘴我们把proto MW和high-z compaction联系起来，给早期星系形成提供了新的观测+模拟限制。
A central goal of Galactic archaeology is to understand the physical processes that governed the earliest phases of the Milky Way's (MW) formation. Ancient stars that retain the chemical and dynamical imprints of their birth environments are ideal tracers. Although precise stellar age measurements remain challenging, especially for old stars, stellar metallicity offers a valuable proxy, since the most metal-poor stars are statistically among the oldest.

In the hierarchical galaxy formation scenario, the inner $\sim5$ kpc region of the MW is expected to harbor most of the oldest metal-poor stars \citep[e.g.,][]{El-Badry2018}, consistent with early hints from the HK Survey based on age gradients of halo blue horizontal-branch stars \citep{Beers1985,Preston1991}. Recent observations \citep{BK22,Conroy2022,Rix2022} confirm this prediction, revealing a centrally concentrated, kinematically hot component of metal-poor stars often referred to as the ``proto-Galaxy'' or ``Aurora.'' In addition, orbit-space analyses of inner-Galaxy metal-poor stars have identified distinct substructures such as Shiva and Shakti, interpreted as possible early massive fragments that coalesced at high redshift (perhaps about 12 Gyr ago) \citep{MalhanRix2024}.

The proto-Galaxy is pivotal for reconstructing the assembly history of the inner halo and the Galactic disk system. Combining Gaia astrometry \citep{Gaia2023} with large spectroscopic surveys such as LAMOST \citep{Zhao2012LAMOST} and APOGEE \citep{apogee}, recent studies have identified that the chaotic proto-Galaxy ([Fe/H]$\lesssim -1.3$) represents the first stage of the three-phase MW disk assembly \citep{BK22,Conroy2022,Rix2022,Chandra2024}. This is followed by the spin-up phase ($-1.3\lesssim$[Fe/H]$\lesssim-0.9$) of the thick disk, with a rapid increase in median rotation velocity and a dynamical transition from a dispersion-dominated to a rotation-dominated regime. Finally, the system enters the cool-down phase to form a thin disk. However, most existing proto-Galaxy spectroscopic samples are limited to $\mathrm{[Fe/H]} \gtrsim -2.0$, leaving the first 1-2 Gyr of the MW's formation history largely unconstrained.

Thanks to the narrow/medium-band large-scale photometric surveys such as SMSS \citep{Wolf2018}, SAGES \citep{Zheng2018,Fan2023}, and J/S-PLUS \citep{Cenarro2019,Mendes2019}, it is now possible to obtain large, all-sky samples of metal-poor stars with reliable photometric metallicities as low as $\mathrm{[Fe/H]} \approx -3.5$, along with accurate photometric distance estimates \citep{Huang2022,Huang2023,Huang2024,Huang2025,Huang2026}. In this work, we use these datasets to trace the spatial distribution, the metallicity distribution, and the kinematics of the most metal-poor component of the proto-Galaxy.

So far, the physical origin of the proto-Galaxy remains poorly understood. \citet{BK22} associate the Aurora component with cold, filamentary gas accretion and an irregular stellar distribution. Other scenarios include an early accretion event \citep{Horta2021} and a possible significant contribution from disrupted globular clusters \citep{BK23}. \citet{Chen2025} suggest that the proto-Galaxy may have undergone starburst events. These explanations highlight different aspects of early galaxy assembly, yet a coherent theoretical picture that unifies them is still lacking. A key open question is therefore: What physical mechanism governed the formation of the proto-Galaxy, and how can the diverse observational signatures be interpreted within a consistent framework?

The Letter is organized as follows. Section~\ref{sec:data} describes the observational data. Section~\ref{sec:results} presents three sets of observational constraints that map the proto-Galaxy. Section~\ref{discussion} discusses the possible formation mechanism of the proto-Galaxy and the in situ fraction of our sample. We conclude in Section~\ref{sec:conclusion}.

\section{Data: Merged Photometric Survey Catalogs}\label{sec:data}

\begin{figure*}
	\centering
	\includegraphics[width = \linewidth]{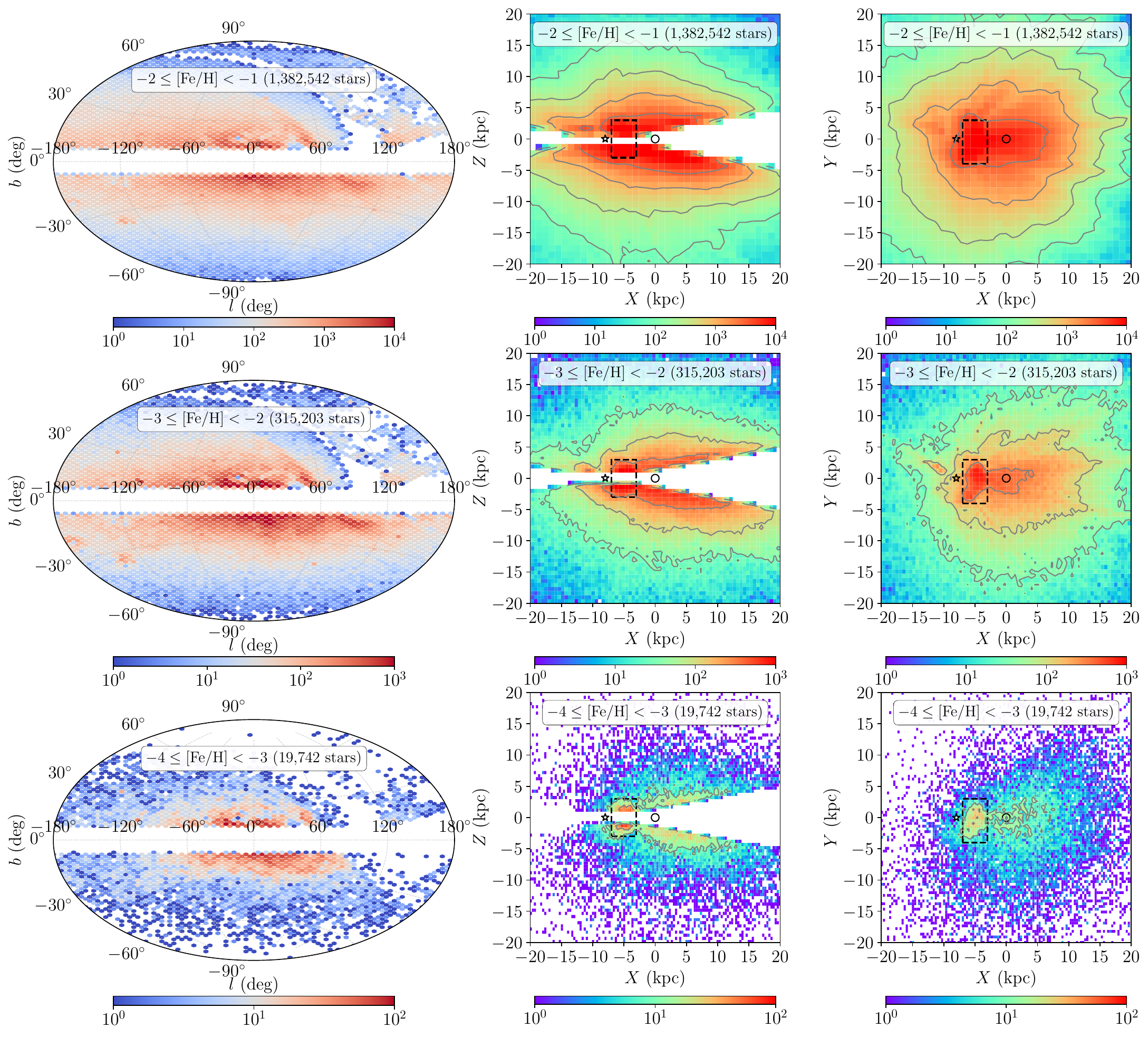}
	\caption{Spatial number-density distributions of the final observed sample in three metallicity bins. The three rows of panels correspond to different metallicity bins: $-2.0 \le \mathrm{[Fe/H]} < -1.0$ (top), $-3.0 \le \mathrm{[Fe/H]} < -2.0$ (middle), and $-4.0 \le \mathrm{[Fe/H]} < -3.0$ (bottom). The first column shows the all-sky distribution in Galactic coordinates $(l, b)$, and the second and third columns present the distribution in Galactocentric Cartesian coordinates $(X, Z)$ and $(X, Y)$ for the final sample, respectively. The open circles and open star symbols in the second and third columns indicate the positions of the Galactic center and the Sun, respectively. The black dashed boxes in the second and third columns mark the region $-7<X<-3\,\mathrm{kpc}$, $-4<Y<3\,\mathrm{kpc}$, and $-3<Z<3\,\mathrm{kpc}$, highlighting the inner-Galaxy overdensity where metal-poor stars with disklike orbits make a significant contribution. In the second and third columns, both axes are divided into $500\times500$ bins. Note that the maximum values of color bars vary across rows: $10^4$ for the top row, $10^3$ for the middle row, and $10^2$ for the bottom row. The number-density contour levels are $[10^{2.0},10^{2.5},10^{3.0},10^{3.5}]$ for the top row, $[10^{1.5},10^{2.0},10^{2.5}]$ for the middle row, and $[10,10^{1.5}]$ for the bottom row. This figure illustrates that, in the inner Galaxy, both a centrally concentrated, flattened spheroidal proto-Galaxy component and a population of metal-poor stars on disklike orbits are present.}
	\centering
	\label{fig:lb_XZ_Au18}
\end{figure*}

We combine the stars classified as giants from \citet{Huang2022,Huang2023,Huang2024,Huang2026} with photometrically estimated stellar metallicities based on the stellar colors of SMSS DR4 \citep[][the updated version of Huang et al. 2022]{Huang2025}, SAGES DR1, J-PLUS DR3, S-PLUS DR4, and Gaia EDR3 \citep{Gaia2021}. Calibrated on training sets with spectroscopic measurements from previous high-, medium-, and low-resolution surveys, the photometric metallicities reach down to [Fe/H]$\sim-3.5$, with typical uncertainties below 0.40 dex. For stars with reliable Gaia EDR3 parallax measurements, distances are estimated by \citet{Bailer-Jones2021} directly. Distances of the remaining stars are measured using empirical color-magnitude fiducials. 

We first select giant stars with Gaia DR3 \citep{Gaia2023} \texttt{ruwe} $<1.4$ from the four photometric catalogs to exclude sources with unreliable astrometry, and retain only stars with available distance estimates. For J-PLUS and S-PLUS giants, we require \texttt{flg}$_{\text{[Fe/H]}} > 0.85$ to ensure the robustness of the metallicity estimates\footnote{\texttt{flg}$_{\text{[Fe/H]}}$ quantifies the reliability of the photometric metallicity, ranging from 0 to 1, with higher values indicating better quality.}. The J-PLUS and S-PLUS catalogs are then merged, keeping only the one with the highest \texttt{flg}$_{\text{[Fe/H]}}$ for stars with multiple measurements.

For SMSS and SAGES, we prioritize [Fe/H] estimates derived from the $v$ band (see the Appendix of \citealt{Hong2024}), followed by those based on the $u$ band. The two catalogs are merged by removing duplicates, keeping the entry with the smallest [Fe/H] uncertainty, err$_{\text{[Fe/H]}}$. Finally, we merge the combined J-PLUS/S-PLUS and SMSS/SAGES samples, giving priority to J-PLUS/S-PLUS entries for duplicated stars.

We compute Galactocentric positions for all stars in the merged photometric sample (J-PLUS/S-PLUS/SMSS/SAGES), and derive full space velocities only for the subset with measured radial velocities (RVs), where the RVs are adopted from large-scale spectroscopic surveys by \citet{Huang2022,Huang2023,Huang2024,Huang2026}. We take the Sun's Galactocentric distance as 8.122 kpc \citep{GRAVITY2018}  and the vertical height above the Galactic plane as 20.8 pc \citep{Bennett2019}. The local standard of rest (LSR) velocity is adopted as 234.04 km\,s$^{-1}$ \citep{Zhou2023}, and the peculiar velocity of the Sun is $(U_{\odot},V_{\odot},W_{\odot})=(11.69,10.16,7.67)$ km\,s$^{-1}$ \citep{Wang2021}. We use a right-hand Galactocentric Cartesian coordinate system $(X,Y,Z)$, where the Sun is at $(X_{\odot},Y_{\odot},Z_{\odot})=(-8.122,0.0,0.0208)$ kpc.

The merged sample is further refined with the following cuts: (1) Galactic latitude $|b| > 10^\circ$ and vertical height $|Z| > 1$ kpc, to exclude regions with poor catalog completeness and also avoid high reddening regions; (2) [Fe/H] uncertainty $\mathrm{err}_{\mathrm{[Fe/H]}} < 1.0$ dex; (3) relative distance uncertainty err$_{d}/d<0.3$; (4) the value\footnote{The $E(B-V)$ value of \citealt{SFD1998} is corrected for a 14\% systematic overestimate (e.g., \citealt{Schlafly2011,Yuan2013}).} of $E(B-V)$ from the extinction map of \citet{SFD1998} less than 0.8, to only retain stars with reliable reddening corrections. The final sample contains 5,095,676 giant stars, including 1,717,610 stars with [Fe/H] $<-1.0$.

We derive the Galactocentric distances of our final sample stars as $r_{\mathrm{gc}}=\sqrt{X^2+Y^2+Z^2}$. For those with RV measurements, we calculate the Galactocentric Cartesian velocities $(V_X,V_Y,V_Z)$ and azimuthal velocities in spherical coordinates $(r,\theta,\phi)$ as $V_{\phi}=V_X \sin\phi - V_Y \cos\phi$. Prograde stars have $V_{\phi}>0$. We define the vertical angular momentum component as $L_Z=R\times V_{\phi}$, where $R=\sqrt{X^2+Y^2}$ is the Galactocentric cylindrical radius. Orbital parameters ($r_{\mathrm{apo}}$,$r_{\mathrm{peri}}$, $Z_{\mathrm{max}}$ and actions) are calculated by \texttt{AGAMA} \citep{Vasiliev2019} with the \texttt{MWPotential2014} potential \citep{Bovy2015}. We define the eccentricity as $e=(r_{\mathrm{apo}}-r_{\mathrm{peri}})/(r_{\mathrm{apo}}+r_{\mathrm{peri}})$. We perform 100 Monte Carlo realizations by resampling the distance, radial velocity, and proper motions according to their quoted Gaussian uncertainties, and adopt the mean and standard deviation of the resulting distributions as the inferred values and uncertainties of the derived kinematic and orbital parameters, respectively.

\section{Results}\label{sec:results}
\begin{figure*}
	\centering
	\includegraphics[width = \linewidth]{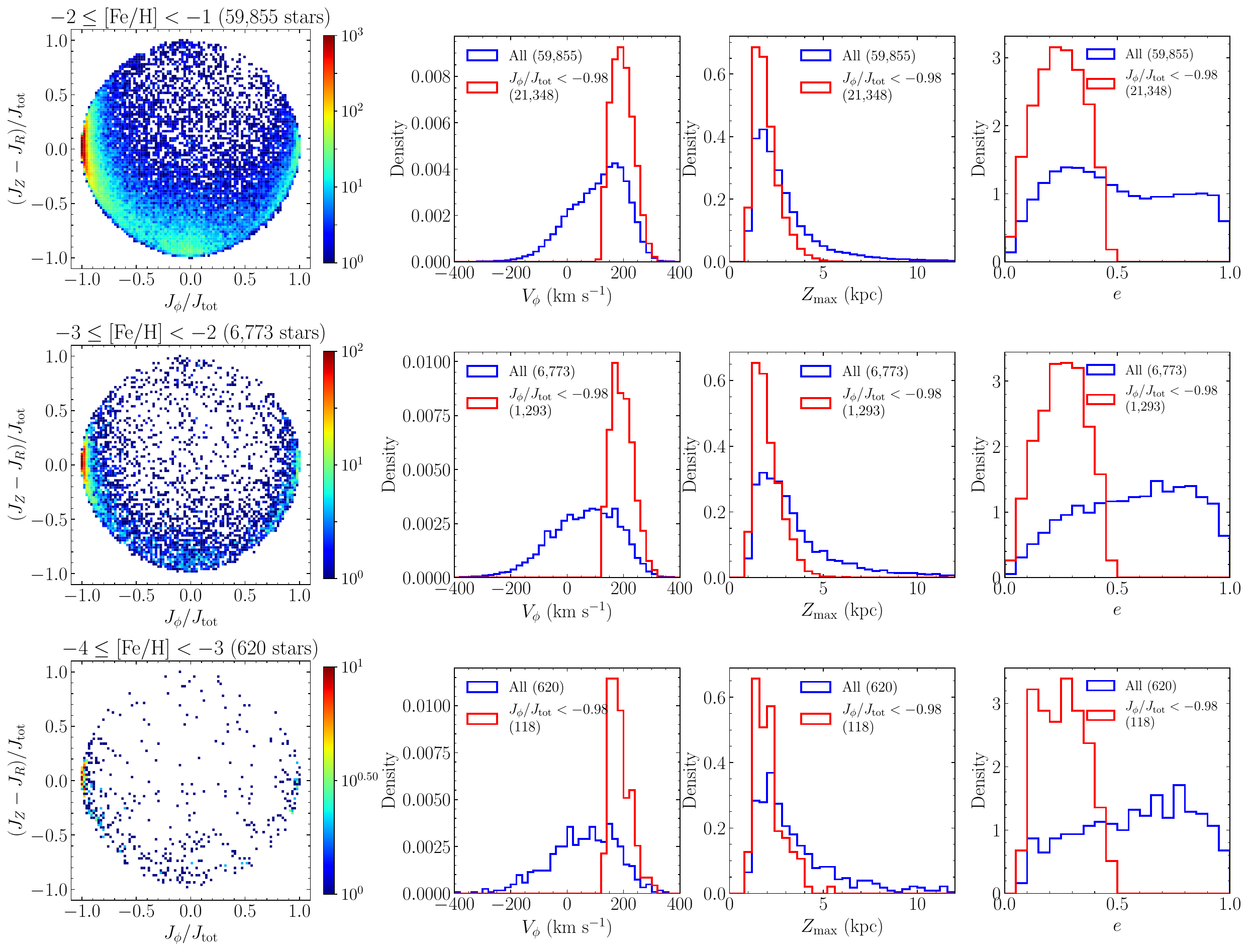}
\caption{Kinematic and orbital-parameter distributions for stars within the overdense inner-Galaxy region highlighted in Figure~\ref{fig:lb_XZ_Au18}, defined by $-7<X<-3\,\mathrm{kpc}$, $-4<Y<3\,\mathrm{kpc}$, and $-3<Z<3\,\mathrm{kpc}$. The three rows correspond to metallicity bins of $-2.0 \le \mathrm{[Fe/H]} < -1.0$ (top), $-3.0 \le \mathrm{[Fe/H]} < -2.0$ (middle), and $-4.0 \le \mathrm{[Fe/H]} < -3.0$ (bottom). The first column shows the stellar number-density distribution in the action diamond space, with the horizontal axis $J_\phi/J_{\rm tot}$ and the vertical axis $(J_Z-J_R)/J_{\rm tot}$, where $J_{\phi}=L_Z$ and $J_{\rm tot}=\sqrt{J_\phi^2+J_Z^2+J_R^2}$. The second to fourth columns show histograms of $V_\phi$, $Z_{\max}$, and eccentricity $e$, respectively. In each metallicity bin, the blue histograms include all stars in the selected region, while the red histograms show the subset with $J_\phi/J_{\rm tot}<-0.98$, following the cut adopted by \citet{Mardini2022} for the ``Atari disk.'' Across all metallicity bins, stars satisfying $J_\phi/J_{\rm tot}<-0.98$ exhibit a coherent prograde rotation with $V_\phi\sim200\,\mathrm{km\,s^{-1}}$, predominantly have $Z_{\max}<3$ kpc and $e<0.5$, and therefore occupy disklike orbits, whereas the remaining stars are kinematically hotter and more halolike. This demonstrates that the overdense region in Figure~\ref{fig:lb_XZ_Au18} contains both metal-poor stars on disklike orbits and a proto-Galaxy/halolike component, with the disklike population dominating the enhancement and the halolike population contributing mainly as a diffuse background.}
	\centering
	\label{fig:overdensity_main}
\end{figure*}

\subsection{The Spatial Distribution of Metal-poor Stars}
We present the currently largest all-sky stellar sample mapping the 3D distribution of metal-poor stars in the inner Galaxy. Figure~\ref{fig:lb_XZ_Au18} shows the spatial density of the final sample in three metallicity bins: $-2.0 \le \mathrm{[Fe/H]} < -1.0$, $-3.0 \le \mathrm{[Fe/H]} < -2.0$, and $-4.0 \le \mathrm{[Fe/H]} < -3.0$. As revealed by the stellar number-density contours in Figure~\ref{fig:lb_XZ_Au18}, stars in all three metallicity bins exhibit a centrally concentrated, flattened spheroidal structure extending to $\sim15$ kpc in Galactocentric radius, with $r_{\mathrm{gc}}$ enclosing 68\% of the stars being 12.2, 17.9, and 15.1 kpc, respectively. This confirms that stars with $\mathrm{[Fe/H]}$ as low as $\sim-3.5$ are preferentially concentrated toward the Galactic center, in agreement with the observational findings of \citet{Rix2022} and the simulation results of \citet{BK22}.

In addition to the global centrally concentrated component, the $(X,Z)$ and $(X,Y)$ maps show a prominent localized overdensity near $X\sim-5$ kpc in all three metallicity bins, which is clearer in the $-3\le\mathrm{[Fe/H]}<-2$ and $-4\le\mathrm{[Fe/H]}<-3$ bins. We manually define an inner-Galaxy region $-7<X<-3\,\mathrm{kpc}$, $-4<Y<3\,\mathrm{kpc}$, and $-3<Z<3\,\mathrm{kpc}$, guided by the high-density features consistently seen across the three metallicity bins in the $(X,Z)$ and $(X,Y)$ projections. This density enhancement is expected to be contributed primarily by metal-poor stars on disklike orbits---a population often discussed in the literature using partially overlapping terms such as the metal-weak thick disk \citep{Morrison1990,Beers2002,Naidu2020}, the ``Atari disk'' \citep{Mardini2022}, or a very metal-poor (VMP; $\mathrm{[Fe/H]}  < -2.0$) disk \citep{Cordoni2021} component. The spatial distribution of the overdense region is also consistent with the literature distributions of metal-poor stars on disklike orbits \citep{An2020,Mardini2022}. In this work, we do not attempt to uniquely label this population; rather, our key empirical point is that metal-poor stars on disklike orbits provide the dominant contribution to the enhancement in this inner-Galaxy region with $-4\le\mathrm{[Fe/H]}<-1$.

The apparent morphology of this overdensity is shaped by the joint impact of our geometric cuts and the survey selection. In the overdense region, the vertical selection is primarily controlled by the $|Z|>1$ kpc requirement. By contrast, in regions around $X\sim0$ and $Y\sim\pm5$ kpc, as well as on the $X>0$ side, the $|b|>10^\circ$ cut becomes the dominant limiter on the accessible $|Z|$ range and preferentially removes stars that would otherwise satisfy $|Z|>1$ kpc, thereby suppressing any comparable enhancement. In addition, on the far side of the Galactic center, the coverage is further reduced by severe extinction, making analogous structures harder to detect. Overall, while the detailed shape and contrast of the overdensity are influenced by these selection effects, orbital diagnostics provide an independent way to assess which populations dominate the enhancement within the selected region.

Figure~\ref{fig:overdensity_main} summarizes the kinematic and orbital properties of stars in the overdense inner-Galaxy region highlighted in Figure~\ref{fig:lb_XZ_Au18}. In the first column, the stellar number density in the action diamond space peaks near $J_\phi/J_{\rm tot}\approx-1$ in all three metal-poor bins. Motivated by this concentration, we adopt an extreme $J_\phi/J_{\rm tot}<-0.98$ cut following \citet{Mardini2022}, who originally introduced this selection to isolate the strongly prograde ``Atari disk'' population, and use it here to examine the orbital properties of the most strongly prograde stars in the overdensity region. The selected stars show coherent prograde rotation and predominantly disklike orbits (low $Z_{\max}$ and low eccentricity), while the remaining stars are kinematically hotter and more halolike. Together, these diagnostics indicate that the overdensity is dominated by the strongly prograde, disklike population, whose projected prominence is largely shaped by the combined $|b|$ and $|Z|$ cuts that define the accessible survey volume, while the kinematically hotter stars contribute more broadly as a diffuse background. The region may also host additional subpopulations that are not explicitly separated here, and we do not attempt a more detailed decomposition in this work; in this sense, the background component can be regarded as the proto-Galaxy/halolike population. The fact that we recover both a metal-poor disklike population and a kinematically hot, proto-Galaxy/halolike population within the same photometric-metallicity-selected sample also provides an internal consistency check that our photometric $\mathrm{[Fe/H]}$ estimates are sufficiently reliable for separating broad populations in the inner Galaxy.

\subsection{The Metallicity Distribution Function of the Inner Galaxy}\label{subsec:MDF}
\begin{figure}
	\centering
	\includegraphics[width = \linewidth]{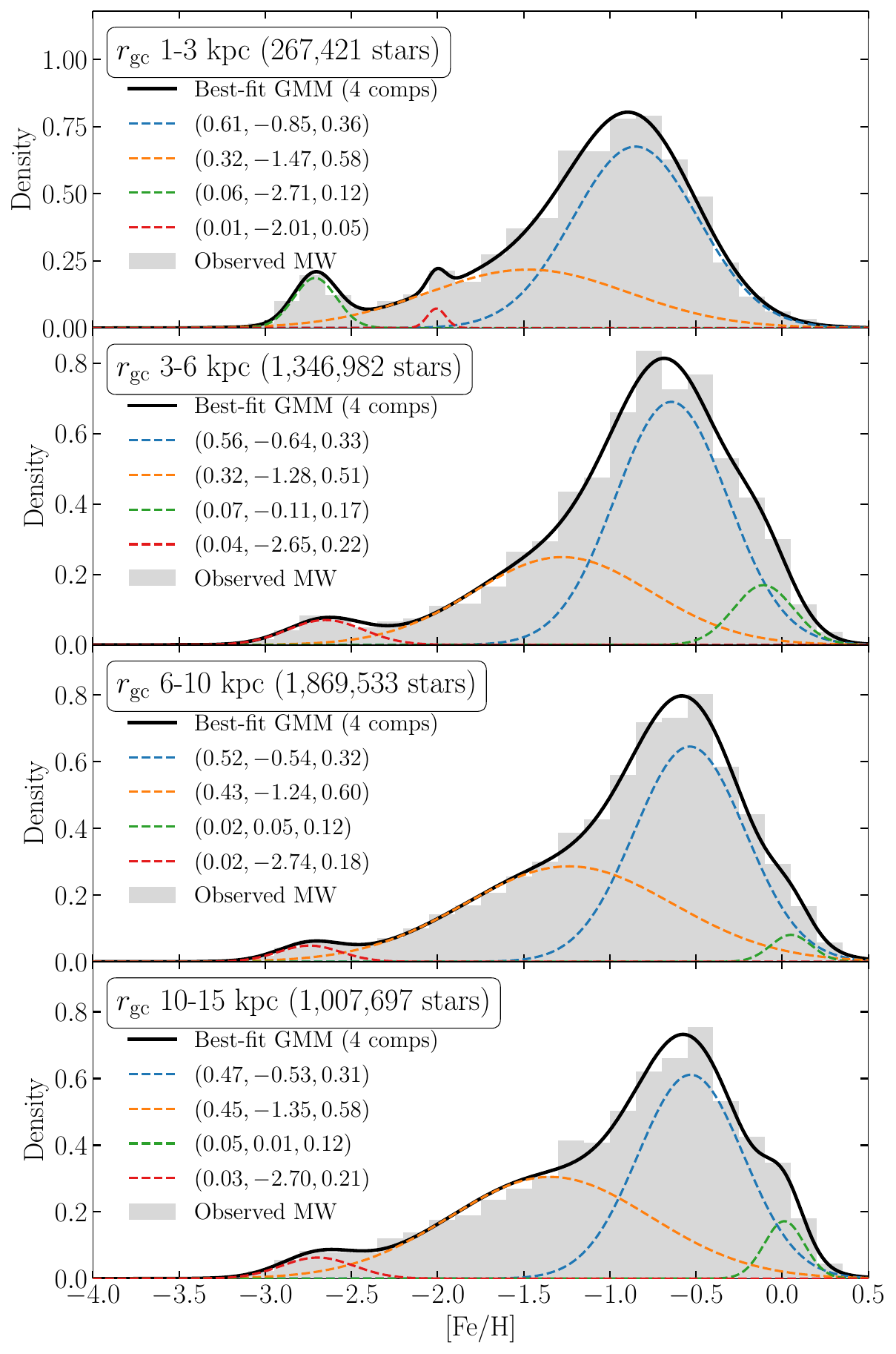}
	\caption{MDFs of the final sample within $r_{\mathrm{gc}}<15$ kpc and the Gaussian mixture model (GMM) fits. Different panels correspond to different $r_{\mathrm{gc}}$ bins, with the number of stars in each bin shown in the top-left corner. The gray histograms are the MDFs of the observed MW stars corrected by the weights derived from the selection function. The optimal number of GMM components in each bin is determined by the Bayesian information criterion (BIC), which favors four components for all bins. Colored dashed curves represent the individual GMM components, and the black solid curve shows their sum. For each component, the corresponding weight, mean, and standard deviation are indicated as $(w,\mu,\sigma)$. The VMP component is present throughout the inner 15 kpc and is relatively prominent in the 1-3 kpc bin.}
	\centering
	\label{fig:MDF_GMM}
\end{figure}

Figure~\ref{fig:MDF_GMM} shows the metallicity distribution function (MDF) of the observed MW stars within a Galactocentric radius of $r_{\mathrm{gc}}<15$ kpc and the Gaussian mixture model (GMM) fits. The observed MDF histograms have been corrected for selection effects relative to Gaia DR3 using color-magnitude diagrams, following \citet{Castro2023}. We start by selecting all stars (both giants and dwarfs) from \citet[][]{Huang2022,Huang2023,Huang2024,Huang2026} using the same procedure described in Section~\ref{sec:data} for obtaining the final sample. The resulting sample is denoted as $T$. Assuming that the selection of stars from Gaia DR3 into the photometric catalogs, and from these catalogs into $T$, does not distinguish between giants and dwarfs, the selection function of our final sample relative to Gaia DR3 can be written as \citep{Castro2023}
\begin{equation}
	S(l, b, G, B P-R P)=\frac{k+1}{n+2}\,,
\end{equation}
where $k$ is the number of stars from sample $T$ in a given $(l,b,G,BP\!-\!RP)$ cell, and $n$ is the number of Gaia DR3 stars in the corresponding cell. We adopt the resolution of HEALPix level4 for sky binning, and bin widths of $0.5$ mag in both $G$ and $BP\!-\!RP$. When constructing the MDF histograms in Figure~\ref{fig:MDF_GMM}, each star is weighted by $S^{-1}$ to account for selection effects. We emphasize that this correction is intended as a first-order approximation, and it may not capture all survey-specific systematics.

We perform the GMM fits using the \texttt{bilby} library \citep{bilby} with the \texttt{nessai} sampler, exploring models with two, three, and four components, and adopt four components for each $r_{\mathrm{gc}}$ bin based on minimization of the Bayesian information criterion (BIC).

Figure~\ref{fig:MDF_GMM} demonstrates that a VMP component at $\mathrm{[Fe/H]}\sim -2.7$ is present across all $r_{\mathrm{gc}}$ bins, and is particularly prominent in the innermost Galaxy, contributing up to $\sim6$\% of the stellar population in the 1--3 kpc bin. This clear VMP peak in the 1--3 kpc bin is one of the most striking features revealed by our large, all-sky inner-Galaxy sample. We caution, however, that the existence and prominence of this feature may be sensitive to the selection correction and other systematics, and will require confirmation with independent datasets and follow-up spectroscopy. A possible physical interpretation is discussed in Section~\ref{subsec:compaction_discussion}, where we note that a prominent VMP component spatially concentrated within the central few kiloparsecs is qualitatively consistent with an early, rapid build-up phase in the proto-Galaxy, as envisioned in high-$z$ compaction/``blue nugget'' (BN) scenarios.

\begin{figure}
	\centering
	\includegraphics[width = \linewidth]{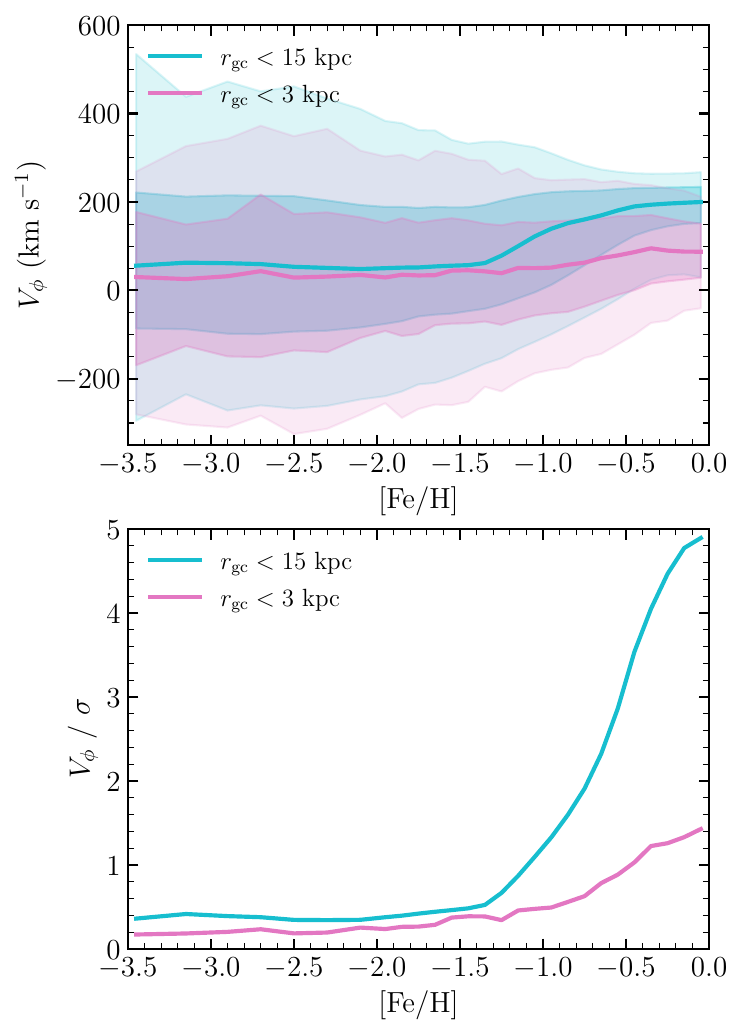}
	\caption{Kinematic trends as functions of metallicity. Top panel: galactocentric rotation velocity $V_{\phi}$ versus [Fe/H] for stars within $r_{\mathrm{gc}}<15$ kpc (blue) and $r_{\mathrm{gc}}<3$ kpc (orchid) of the final sample. Solid lines show the median $V_{\phi}$, with dark- and light-shaded regions indicating the 16$^{\rm th}$-84$^{\rm th}$ and 2.5$^{\rm th}$-97.5$^{\rm th}$ interpercentile ranges, respectively. Bottom panel: rotational support, $V_{\phi}/\sigma$, where $V_{\phi}$ median is adopted and $\sigma$ is estimated as half the 16$^{\rm th}$-84$^{\rm th}$ interpercentile range of $V_{\phi}$ in each [Fe/H] bin. The spatial selection is identical to the top panel.}
	\centering
	\label{fig:Vphi_FEH_sigma}
\end{figure}

\subsection{Kinematics}\label{subsec:kinematics}
%图的构思：竖着三张子图，第一张是Vphi-[Fe/H]，第二张是(Vphi/sigma)-[Fe/H]，第三张是Au18的Age-[Fe/H]。
%之后的main sequence图，可以仿照Semenov把一些关键节点的edge on图像画上。

Figure~\ref{fig:Vphi_FEH_sigma} shows how the median azimuthal velocity $V_\phi$ (top) and rotational support $V_\phi/\sigma$ (bottom) vary with metallicity for stars from the final sample, comparing an inner-Galaxy sample with $r_{\rm gc}<15$ kpc to a more central subsample with $r_{\rm gc}<3$ kpc. For $r_{\rm gc}<15$ kpc, both $V_\phi$ and $V_\phi/\sigma$ remain low and nearly flat at $\mathrm{[Fe/H]}\lesssim -1.4$ ($V_\phi\sim 50~\mathrm{km\,s^{-1}}$ and $V_\phi/\sigma\sim 0.4$), but rise sharply for $\mathrm{[Fe/H]}\gtrsim -1.4$, reaching $V_\phi\sim 200~\mathrm{km\,s^{-1}}$ by $\mathrm{[Fe/H]}\sim -0.5$ and $V_\phi/\sigma\sim 5$ near solar metallicity. This rapid increase is consistent with the commonly discussed spin-up feature \citep{BK22}, in which progressively more metal-rich populations exhibit increasingly strong rotational support as the Galactic disk becomes established.  

In contrast, the $r_{\rm gc}<3$ kpc subsample shows a much weaker metallicity dependence: $V_\phi$ rises only gradually above $\mathrm{[Fe/H]}\approx -1.4$ and remains $\lesssim 100~\mathrm{km\,s^{-1}}$ even at $\mathrm{[Fe/H]}\sim 0$, while $V_\phi/\sigma$ increases slowly from $\sim0.2$ to $\sim1.4$ without a pronounced spin-up knee. A natural interpretation is that, at such small radii and at $|Z|>1$ kpc, the population remains strongly influenced by a kinematically hot proto-Galaxy component on halolike orbits; increasing metallicity then primarily changes the relative mixture of hot and rotating populations rather than producing a wholesale transition to a cold, rapidly rotating disk.  

\section{Discussion}\label{discussion}
\subsection{In situ or accreted?}\label{subsec:insitu_accreted}
\begin{figure}
	\centering
	\includegraphics[width = \linewidth]{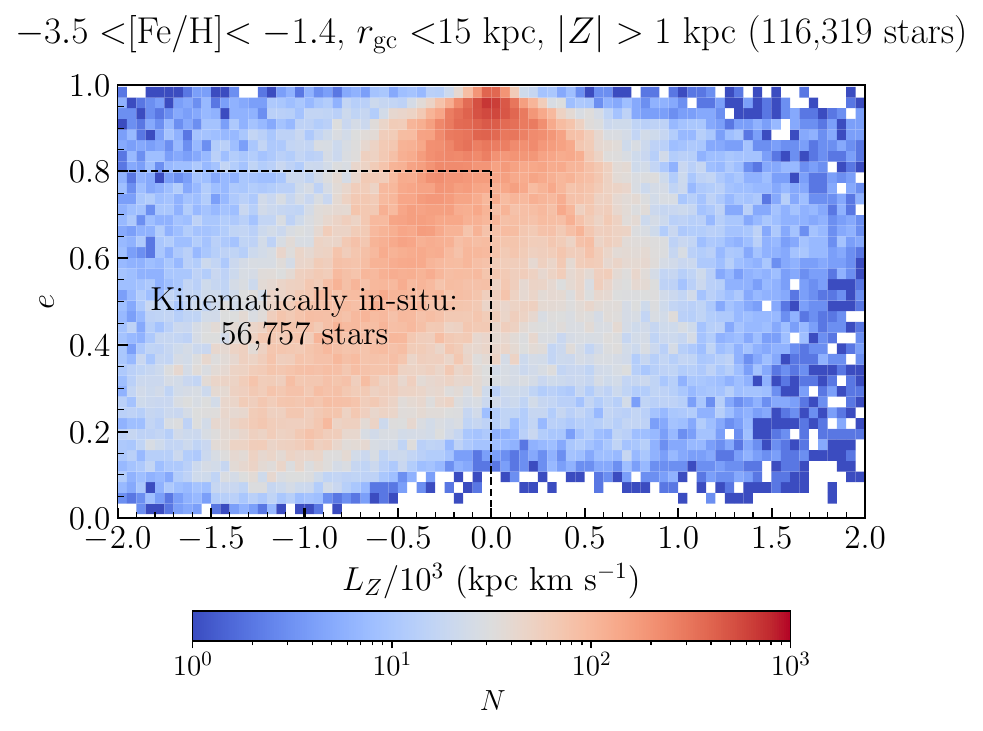}
	\caption{
Stellar number-density distribution of the proto-Galaxy sample in the plane of orbital eccentricity $e$ versus angular momentum along the $Z$-axis, $L_Z$. The sample includes 116,319 stars from the final sample with $-3.5 < \mathrm{[Fe/H]} < -1.4$, $r_{\mathrm{gc}} < 15$ kpc, and $|Z| > 1$ kpc, selected to represent the proto-Galaxy population. The upper metallicity threshold is motivated by the spin-up metallicity inferred from Figure~\ref{fig:Vphi_FEH_sigma}. The black dashed box shows the kinematically in situ region defined by \citet{Conroy2022} ($e < 0.8$ and $L_Z > 0$ kpc\,km\,s$^{-1}$), which contains 56,757 stars, i.e., a conservative lower limit of 48.8\% of the total proto-Galaxy sample shown in this figure.
}
	\centering
	\label{fig:FEH_rgc_e_Lz}
\end{figure}

It remains unclear whether the most metal-poor stars in the inner Galaxy were born in the main progenitor of the MW or were accreted. As discussed in \citet{Rix2022}, the in situ and accreted stars are hardly separable in the chaotic early Galaxy by spatial distribution and chemical composition. Chemical abundances like [Al/Fe] \citep{BK22} can separate these populations for stars with [Fe/H] $\gtrsim -1.2$, but this diagnostic fails at lower metallicities, where both in situ and accreted stars occupy overlapping regions in chemical space \citep[e.g.,][]{Horta2021,Conroy2022}.

Figure~\ref{fig:FEH_rgc_e_Lz} presents the orbital properties of our proto-Galaxy sample ($-3.5 < \mathrm{[Fe/H]} < -1.4$, $r_{\mathrm{gc}} < 15$ kpc, and $|Z| > 1$ kpc) in the $L_Z$--eccentricity space for 116,319 stars with complete 6D kinematics. The upper metallicity threshold $\mathrm{[Fe/H]}=-1.4$ is motivated by the spin-up metallicity of stars within $r_{\mathrm{gc}}<15$ kpc inferred from Figure~\ref{fig:Vphi_FEH_sigma}. Following \citet{Conroy2022}, we adopt a kinematic criterion for in situ stars as those with $e < 0.8$ and $L_Z > 0$ kpc\,km\,s$^{-1}$ (dashed box). This cut yields a relatively pure but incomplete selection of in situ stars, and therefore the resulting fraction should be interpreted as a conservative lower limit, since it excludes in situ stars on high-eccentricity or retrograde orbits. With this criterion, we identify 56,757 kinematically in situ stars, corresponding to a lower limit of 48.8\% of the proto-Galaxy sample in Figure~\ref{fig:FEH_rgc_e_Lz}.

Our results suggest that at least half of our proto-Galaxy sample likely formed in situ. Given the ambiguity between in situ and accreted stars in the early MW, we retain the full final sample without kinematic filtering in this work. This approach allows us to trace the complete chemodynamical signatures of most metal-poor stars in the inner Galaxy.

\subsection{A possible link to high-$z$ compaction/``blue-nugget'' phases}\label{subsec:compaction_discussion}
Over the past decade, both observations and simulations have revealed that the most compact star-forming galaxies (SFGs) at $z \gtrsim 2$ are likely shaped by highly dissipative gas-rich compaction processes \citep[][]{Dekel2014,Tacchella2016,Lapiner2023}. These compaction events, triggered by so-called 
``wet'' (gas-rich) mergers, counterrotating cold streams, or violent disk instabilities, lead to epochs of gas condensation and subsequent central starbursts, creating young, star-forming central regions with high stellar densities and rapid gas depletion, dubbed BN phases. The BN phases, occurring around a characteristic stellar mass of $10^{9.5-10}\,M_\odot$, trigger central quenching and mark drastic transformations in the structural, kinematic, and compositional properties of galaxies.

In this framework, compaction can drive an oscillatory evolution around the star-forming main sequence: galaxies rise above the main-sequence ridge during compaction/BN phases, then quickly move below it as the central gas is consumed or expelled \citep[][]{Tacchella2016}. At high redshift, when halo masses are relatively low, quenching may be temporary; subsequent gas replenishment can enable repeating episodes of compaction. While our Letter is intentionally focused on the observational constraints and does not attempt to establish a unique formation pathway, this picture provides a useful context for discussing whether early, centrally concentrated low-metallicity components in MW-like systems could plausibly arise from one or more such dissipative build-up episodes. It is worth noting that our data reveal a potentially suggestive chemical signature (Section~\ref{subsec:MDF}): the metallicity distribution shows a distinct VMP component around $\mathrm{[Fe/H]}\sim-2.7$ that is present at all radii of $r_{\mathrm{gc}}<15$ kpc but becomes most prominent in the inner $1$--$3$~kpc, contributing up to $\sim6$\% of the stars in that radial bin.

If this centrally concentrated VMP component is confirmed to be robust against remaining selection and distance systematics, one possible interpretation is that it reflects one or more early, rapid build-up episodes in the main progenitor of the MW. Qualitatively, this scenario could therefore accommodate (i) a centrally enhanced fraction of VMP stars and (ii) the predominantly dispersion-supported kinematics of the low-metallicity population, which would be qualitatively consistent with a formation channel involving multidirectional, gas-rich inflows (e.g., wet mergers and misaligned/counterrotating streams) that efficiently mix and partially cancel angular momentum.

At the same time, our present observations do not uniquely require a compaction-driven origin. A centrally enhanced VMP component could also arise from early accretion of low-mass substructures whose debris is deposited preferentially at small radii, from a mixture of in situ and accreted populations in the chaotic early Galaxy, or from other dissipative processes unrelated to classic high-$z$ BN phases. Distinguishing among these possibilities will require additional constraints: precise abundance patterns, improved age information where available, and forward modeling of selection effects coupled with comparisons to ensembles of MW analog simulations. In this sense, the inner $1$--$3$~kpc VMP component provides a useful empirical target for future chemodynamical modeling of the proto-Galaxy.

\section{Conclusion}\label{sec:conclusion}

We have constructed the currently largest 3D chemodynamical map of metal-poor giants in the inner Galaxy by merging several narrow/medium-band photometric surveys \citep{Huang2022,Huang2023,Huang2024,Huang2026}. The final sample contains 5,095,676 giant stars, including 1,717,610 stars with $\mathrm{[Fe/H]}<-1$, and reaches $\mathrm{[Fe/H]}\sim-3.5$.

Our main observational results are as follows:
\begin{enumerate}
    \item Global inner-Galaxy structure. In all three metal-poor bins ($-2\le\mathrm{[Fe/H]}<-1$, $-3\le\mathrm{[Fe/H]}<-2$, and $-4\le\mathrm{[Fe/H]}<-3$), the density distribution reveals a centrally concentrated, flattened spheroidal component extending to $r_{\rm gc}\sim15$ kpc.
    
    \item A disk-dominated inner-Galaxy overdensity. Superposed on the global component, we identify a prominent overdensity near $X\sim-5$ kpc. Orbital diagnostics show that this feature is dominated by metal-poor stars on disklike orbits, with a kinematically hot proto-Galaxy/halolike component contributing mainly as a diffuse background.
    
    \item A distinct VMP component in the inner MDF. After correcting for selection effects relative to Gaia DR3, the MDF within $r_{\rm gc}<15$ kpc is well described by four GMM components in each radial bin. A VMP component centered at $\mathrm{[Fe/H]}\sim-2.7$ is present at all radii and is most prominent at 1--3 kpc, where it contributes up to $\sim6\%$ of the stellar population.
    
    \item Metallicity-dependent kinematics and the spin-up transition. For $r_{\rm gc}<15$ kpc, stars with $\mathrm{[Fe/H]}\lesssim-1.4$ show weak net rotation and low rotational support ($V_\phi\sim50~\mathrm{km\,s^{-1}}$ and $V_\phi/\sigma\sim0.4$), followed by a sharp increase in both $V_\phi$ and $V_\phi/\sigma$ at higher metallicity. Within $r_{\rm gc}<3$ kpc, the metallicity dependence is much weaker, and the population remains kinematically hot.
    
    \item A conservative in situ lower limit. For a proto-Galaxy sample defined by $-3.5<\mathrm{[Fe/H]}<-1.4$, $r_{\rm gc}<15$ kpc, and $|Z|>1$ kpc, 48.8\% satisfy a conservative kinematic in situ criterion ($e<0.8$ and $L_Z>0$), implying a lower limit of about half formed in situ.
\end{enumerate}

These results provide new, large-scale empirical constraints on the 3D structure, MDF subcomponents, and metallicity-dependent kinematics of the metal-poor inner Galaxy and its proto-Galaxy population. The centrally enhanced VMP component at 1--3 kpc is a particularly useful observational target for future chemodynamical modeling; as one possible interpretation, it may be qualitatively consistent with early dissipative build-up episodes such as high-$z$ compaction/``BN'' phases. Follow-up high-resolution spectroscopy and improved forward modeling of selection effects will be essential to test the origin of this component.

\section*{acknowledgments}
This work is supported by the National Key R\&D Program of China No. 2024YFA1611903. Y.H. acknowledges the support from the National Natural Science Foundation of China (NSFC grant No. 12422303), the Fundamental Research Funds for the Central Universities (grant Nos. 118900M122, E5EQ3301X2, and E4EQ3301X2), and the National Key R\&D Program of China (grant No. 2023YFA1608303). F.J. acknowledges support by the National Natural Science Foundation of China (NSFC, 12473007). X.-X.X. acknowledges the support from the National Key Research and Development Program of China No. 2024YFA1611902, National Natural Science Foundation of China (NSFC) No. 12588202, CAS Project for Young Scientists in Basic Research grant No. YSBR-062, the Strategic Priority Research Program of Chinese Academy of Sciences grant No. XDB1160102, and the science research grants from the China Manned Space Project with NO. CMS-CSST-2025-A11. T.C.B. acknowledges partial support from grants PHY 14-30152; Physics Frontier Center/JINA Center for the Evolution of the Elements (JINA-CEE), and OISE-1927130; The International Research Network for Nuclear Astrophysics (IReNA), awarded by the US National Science Foundation, and DE-SC0023128; the Center for Nuclear Astrophysics Across Messengers (CeNAM), awarded by the U.S. Department of Energy, Office of Science, Office of Nuclear Physics. C.C. and J.A. are supported by the National Natural Science Foundation of China under grant Nos. 12233001 and 12533004, by the National Key R\&D Program of China under grant No. 2024YFA1611602, by a Shanghai Natural Science Research Grant (24ZR1491200), by the ``111'' project of the Ministry of Education under grant No. B20019, and by the China Manned Space Program with grant Nos. CMS-CSST-2025-A08, CMS-CSST-2025-A09, and CMS-CSST-2025-A11.

\bibliography{sample631}{}
\bibliographystyle{aasjournal}



\end{document}